\documentclass[10pt,aps,prc,floatfix,twocolumn,nofootinbib]{revtex4-1}

\usepackage{amsfonts}
\usepackage{amsmath}
\usepackage{amssymb}
\usepackage{bbold}
\usepackage{bm}
\usepackage{cellspace}
\usepackage{color}
\usepackage{enumerate}
\usepackage{graphicx}
\usepackage{physics} % For bra-ket notation
\usepackage[figuresright]{rotating}
\usepackage{siunitx}
\usepackage[caption=false]{subfig} % For sub-figures
\usepackage{simpler-wick}
\usepackage{xfrac}
\usepackage{hyperref} % For clickable links to sections within table of contents

\graphicspath{{figures/}} % Setting the graphics path

\newcommand{\sgn}[0]{\sigma}

\newcommand{\akdag}[0]{a^{\dagger}}
\newcommand{\ak}[0]{a^{\phantom{\dagger}}}

\newcommand{\Flo}{F^{\rm lo}}
\newcommand{\Fhi}{F^{\rm hi}}
\newcommand{\mystrut}{\rule{0pt}{0.9\normalbaselineskip}}

\newcommand{\kvec}[0]{\mathbf{k}}
\newcommand{\kpvec}[0]{\mathbf{k'}}

\newcommand{\qvec}[0]{\mathbf{q}}

\newcommand{\Qvec}[0]{\mathbf{Q}}

\newcommand{\kF}{k_{\scriptscriptstyle{\text F}}}

\newcommand{\fmi}{\ensuremath{\mbox{fm}^{-1}}}

\newcommand{\nhat}{\widehat{n}} % Momentum distribution
\newcommand{\nhathi}{\nhat_{\infty}} % High RG momentum distribution
\newcommand{\nhatlo}{\nhat_{\lambda}} % Low RG momentum distribution

\newcommand{\tripletS}{\ensuremath{^3\mbox{S}_1}}
\newcommand{\tripletD}{\ensuremath{^3\mbox{D}_1}}

\newcommand{\rvec}[0]{\mathbf{r}}

\newcommand{\chiEFT}{\ensuremath{\chi{\textrm{EFT}}}} % Chiral EFT
\newcommand{\NNLO}{N$^2$LO}

\newcommand{\NNNNLO}{N$^4$LO}

\newcommand{\Uhat}[0]{\widehat U} % No \lambda subscript
\newcommand{\Uhatdag}[0]{\widehat U^{\dagger}}
\newcommand{\Uhatlam}[0]{\Uhat_{\lambda}^{\phantom{\dagger}}}
\newcommand{\Uhatlamdag}[0]{\Uhat_{\lambda}^{\dagger}}
\newcommand{\Ihat}[0]{\widehat I}

\newcommand{\Ohat}[0]{\widehat O} % No \lambda subscript

\begin{document}

%%%%%%%%%%%%%%%%%%%%%%%%%%%%%%%%%%%%%%%%%%%%%%%%%%%%%%%%%%%%%%%%%%%%%%%%%
\title{The quasi-deuteron model at low RG resolution}

\author{A.~J.~Tropiano$^{1}$, S.~K.~Bogner$^{2}$, R.~J.~Furnstahl$^{1}$,
M.~A.~Hisham$^{1}$}

\affiliation{%
$^1$\mbox{Department of Physics, The Ohio State University, Columbus, OH 43210, USA}  \\
$^2$\mbox{Facility for Rare Isotope Beams and Department of Physics and Astronomy,}  \\
\mbox{Michigan State University, East Lansing, MI 48824, USA}
}

\date{\today}

\begin{abstract}

\begin{description} 
    \item[Background] The quasi-deuteron model introduced by Levinger is used to explain cross sections for knocking out high-momentum protons in photo-absorption on nuclei.
    This is within a framework we characterize as exhibiting high renormalization group (RG) resolution.
    Assuming a one-body reaction operator, the nuclear wave function must include two-body short-range correlations (SRCs) with deuteron-like quantum numbers. In [Tropiano et al., Phys. Rev. C \textbf{104}, 034311 (2021)], we showed that SRC physics can be naturally accounted for at \emph{low} RG resolution.
    \item[Purpose] 
    Here we describe the quasi-deuteron model at low RG resolution     
    and determine the Levinger constant, which is proportional to the ratio of nuclear photo-absorption to that for photo-disintegration of a deuteron.
    \item[Method] We extract the Levinger constant based on the ratio of momentum distributions at high relative momentum.
    We compute momentum distributions evolved under similarity RG (SRG) transformations, where the SRC physics is shifted into the operator as a universal two-body term.
    The short-range nature of this operator motivates using local-density approximations with uncorrelated wave functions in evaluating nuclear matrix elements, which greatly simplifies the analysis.
    The operator must be consistently matched to the RG scale and scheme of the interaction for a reliable extraction.
    We apply SRG transformations to different nucleon-nucleon (NN) interactions and use the deuteron wave functions and Weinberg eigenvalues to determine approximate matching scales.
    \item[Results] We predict the Levinger constant for several NN interactions and a wide range of nuclei comparing to experimental extractions.
    \item[Conclusions] The predictions at low RG resolution are in good agreement with experiment when starting with a hard NN interaction and the initial operator.
    Similar agreement is found using soft NN interactions when the additional two-body operator induced by evolution from hard to soft is included.
\end{description}

\end{abstract}

\maketitle

%%%%%%%%%%%%%%%%%%%%%%%%%%%%%%%%%%%%%%%%%%%%%%%%%%%%%%%%%%%%%%%%%%%%%%%%%
\section{Introduction}
\label{sec:introduction}

Changing the renormalization group (RG) resolution is a powerful technique for analyzing nuclear processes.
In this context, the RG resolution is the scale of the largest momentum components in the wave functions of low-energy states.
It should not be confused with the experimental resolution that is set by the kinematics of a process and which does not change under RG evolution.
In Ref.~\cite{Tropiano:2021qgf}, we showed how evolving to low resolution quantitatively accounts for short-range correlation (SRC) physics phenomenology~\cite{Hen:2016kwk,Korover:2014dma,Hen:2014nza,Duer:2018sby,Duer:2018sxh,Schmookler:2019nvf,Cruz-Torres:2020uke,Schmidt:2020kcl,CLAS:2020rue,Arrington:2022sov} with a cleanly interpreted framework that enables simple yet systematically improvable approximations.
More generally, RG evolution enhances scale separation and hence factorization of structure and reaction mechanisms, facilitating the extraction of process-independent quantities and correlations between observables.
Here we show how the quasi-deuteron model fits into this framework.

The quasi-deuteron model was introduced long ago by Levinger to explain the knock-out of high-energy protons in photo-absorption on nuclei at energies of order 100\,MeV~\cite{Heidmann:1950zz,Levinger:1951vp,Levinger:1979jf,Levinger:2002vj}.
In particular, the emitted protons were argued to originate from two-body SRCs with deuteron-like quantum numbers (``quasi-deuterons'') in the nuclear wave function.
The quasi-deuterons are induced by the nucleon-nucleon (NN) interaction.
Only proton-neutron ($pn$) pairs are relevant because the dipole term in the photoelectric effect is expected to dominate at the photon energies considered~\cite{Weiss:2014gua}.
The picture is that the photon is absorbed by a correlated $pn$ pair (the SRC), followed by the emission of the $pn$ pair back-to-back, without any further interaction.

The consequence is a proportionality of the photo-absorption cross section of a nucleus with $Z$ protons and $N$ neutrons, $A = N+Z$, to that for photo-disintegration of the deuteron,
\begin{equation}
    \label{eq:cross_section_ratio}
    \sigma_{A}(E_{\gamma}) = L \frac{NZ}{A} \sigma_d(E_{\gamma})
    ,
\end{equation} 
where $E_{\gamma}$ is the energy of the photon.%
\footnote{We suppress here a factor that accounts for Pauli blocking in the final state~\cite{Levinger:1979jf,Terranova:1989aa,Chadwick:1991zzb}.}
The Levinger constant $L$ is independent of energy as the cross sections have the same energy dependence, dictated by factorization of short-distance physics~\cite{Tropiano:2021qgf}.
In Ref.~\cite{Levinger:1951vp}, Levinger approximated the ratio of cross sections as an energy-independent ratio of squared wave functions,
\begin{equation}
    \label{eq:wave_func_ratio}
    \frac{|\psi_k|^2}{|\psi_d|^2} \approx L \frac{NZ}{A}
    ,
\end{equation} 
where $\psi_k$ and $\psi_d$ are the in-medium $pn$ pair and deuteron wave functions, respectively.
The factor $NZ/A$ follows from scaling by the number of $pn$ pairs, so that $L$ is a dimensionless measure of the density of quasi-deuterons.
Eq.~\eqref{eq:wave_func_ratio} shows that the cross section ratio effectively counts the relative probability of quasi-deuterons in a nucleus.
See Ref.~\cite{Gottfried:1958aa} for a detailed discussion and derivation of the quasi-deuteron model.
Assuming the quasi-deuteron model is a good approximation, $L$ is a ratio of measurable quantities, and is therefore scale and scheme independent.

A modern treatment of SRCs by Weiss and collaborators relates the Levinger constant to so-called ``nuclear contacts'' in a model called the generalized contact formalism (GCF)~\cite{Weiss:2014gua,Weiss:2015mba,Weiss:2015pjw}.
In these references, the $A/d$ photo-absorption cross section ratio is expressed in terms of nuclear contacts, which measure the probability to find two unlike nucleons close to each other.
Therefore, the ratio gives the relative probability of finding SRC pairs in the nucleus.
The many-body wave function $\Psi$ is factorized into an asymptotic pair wave function $\phi_{ij}(\rvec_{ij})$ and $A_{ij}$, which is the regular part of $\Psi$ describing the residual $A-2$ system and depends on the contacts.
The $\phi_{ij}(\rvec_{ij})$ wave function is fixed in the two-body system, and thus cancels in the ratio of cross sections, leaving dependence on the contacts only.

At high relative momentum, the ratio of momentum distributions is given by contacts as well~\cite{Weiss:2015mba}:
\begin{equation}
    \label{eq:gcf_contact_ratio}
    \frac{ F_{pn}(X) }{ n_p(^2\textrm{H}) } \approx \frac{ C_{pn}^{s_2=0}(X)+C_{pn}^{s_2=1}(X) }{ C_{pn}^{s_2=1}(^2\textrm{H}) } ,
\end{equation}
where $X$ represents the nucleus and $C_{pn}^{s_2}$ are the $pn$ contacts with total spin $s_2$.
Using the $pn$ contacts, $L$ is shown to be proportional to the ratio of the $pn$ pair relative momentum distribution $F_{pn}$ and the proton momentum distribution of the deuteron $n_p(^2H)$,
\begin{equation}
    \label{eq:gcf_levinger}
    \frac{ F_{pn}(X) }{ n_p(^2H) } \approx L \frac{NZ}{A}
    .
\end{equation}
Consistent with the quasi-deuteron model, which assumes only two-body contributions, Weiss et al.\ truncate three-body correlations under the assumption that they contribute much less than the two-body correlations~\cite{Weiss:2015pjw}.

The RG offers an alternative analysis that is simple and universal.
The picture that emerges for the considered photo-absorption kinematics is that the single-nucleon reaction operator that dominates at high RG resolution evolves to include a dominant \emph{two-body} operator at low RG resolution.
The quasi-deuteron model is now manifested as this two-body operator that is common to nuclear photo-absorption and deuteron photo-disintegration.
Factorization of reaction and structure (rather than factorization of the many-body wave function as in the GCF) makes clear that the Levinger constant only involves long-distance physics that should be well-treated by simple approximations and amenable to systematic corrections.

In this paper, we extract the Levinger constant using the ratio of momentum distributions at high relative momentum as in Eq.~\eqref{eq:gcf_levinger}.
Our predictions utilize interactions ranging from Argonne v18 (AV18)~\cite{Wiringa:1994wb} to soft \chiEFT\ interactions.
To account for the scale dependence associated with different interactions, we must include an additional induced two-body operator found by applying inverse-SRG transformations of a harder potential.
This is analogous to reaction operators inheriting the RG scale and scheme of the underlying Hamiltonian.
In an exact low RG resolution calculation, every component in the transition matrix element must be SRG-evolved: the initial and final states, and the electromagnetic operator.
Each component would change; however, due to the unitarity of SRG transformations, the matrix element would stay the same preserving the cross section.
In this picture, the induced two-body operator acting on low-momentum nucleons described by an uncorrelated initial state is responsible for ejected high-momentum nucleons.
Refs.~\cite{More:2015tpa,More:2017syr} demonstrate these concepts for deuteron electrodisintegration, though the consequences follow more generally for breakup/knockout reactions.

In Section~\ref{sec:formalism} we provide the necessary formalism for the low-resolution treatment, building on the developments in Ref.~\cite{Tropiano:2021qgf}.    
Results are given in Section~\ref{sec:results} for several nuclei and compared with experimental extractions.
We also examine scale and scheme dependence in extracting $L$ under various NN interactions. 
Section~\ref{sec:summary} has a summary and outlook.

%%%%%%%%%%%%%%%%%%%%%%%%%%%%%%%%%%%%%%%%%%%%%%%%%%%%%%%%%%%%%%%%%%%%%%%%%
\section{Formalism at low RG resolution}
\label{sec:formalism}

SRG transformations when applied to NN potentials decouple momentum scales~\cite{Bogner:2006pc,Bogner:2009bt,Furnstahl:2013oba,Hergert:2016iju}.
A common decoupling scheme is to drive the potential to band-diagonal form in momentum space as a function of the flow parameter $\lambda$, where $\lambda^2$ roughly measures the width of the band-diagonal potential with respect to relative momentum squared.
Unevolved potentials start at $\lambda=\infty$ and are typically evolved to some finite value of $\lambda$ by integrating the flow equation
\begin{equation}
    \label{eq:flow_equation}
    \frac{dV_{\lambda}}{d\lambda} = -\frac{4}{\lambda^5} [\eta_{\lambda},H_{\lambda}]
    ,
\end{equation}
where $H_{\lambda}$ is the evolving Hamiltonian and $\eta_{\lambda}$ is the anti-hermitian SRG generator defined as the commutator $\eta_{\lambda}=[G_{\lambda},H_{\lambda}]$.
Choosing the operator $G_{\lambda}$ specifies a decoupling scheme.
In this paper we set $G_{\lambda}=H^d_{\lambda}$, that is, the diagonal of the evolving Hamiltonian.
Other operators can be evolved either by solving an analogous flow equation to~\eqref{eq:flow_equation} or by applying SRG transformations directly.
In the latter case, one can construct the transformations from the eigenvectors of the initial and evolved Hamiltonians (see Ref.~\cite{Anderson:2010aq,Tropiano:2020zwb} for details on operator evolution).

In this paper we take $\lambda = 1.35\,\fmi$ as a representative low-resolution RG scale for nuclear ground states that sets the dividing line between low and high momentum.
This separation of momentum scales has significant implications when applying consistently evolved operators that probe a high-momentum scale $q$.
Decoupling in the potential leads to suppression of momenta above $\lambda$ in low-energy states, hence low resolution.
For $q \gg \lambda$, high-momentum operator expectation values are factorized into a product of a universal two-body function encompassing the high-momentum dependence, and a state-dependent matrix element sensitive only to low momenta.
This contrasts to factorization in the GCF where the nuclear wave function factorizes into a universal two-body wave function and a contact dependent term.
The details of factorization from the SRG standpoint can be found in Refs.~\cite{Anderson:2010aq,Bogner:2012zm} and a schematic version is presented in Ref.~\cite{Tropiano:2021qgf}.
For clarity, we will use $q$ to denote the high-momentum scale and $k$ for the low-momentum scale.

At low RG resolution we calculate momentum distributions by using SRG transformations to evolve initial operators (truncating three-body and higher contributions) and evaluate nuclear matrix elements under simple approximations.
As in Ref.~\cite{Tropiano:2021qgf}, we will use a local-density approximation (LDA) and average over local Fermi momentum $\kF^\tau(R)$ to evaluate nuclear matrix elements.
For brevity we will only repeat the key points in this section while further detail can be found in~\cite{Tropiano:2021qgf}.

The SRG unitary transformation at flow parameter $\lambda$ has the following form in second quantization:
\begin{align}
    \label{eq:Uschematic}
    \Uhat_\lambda = \Ihat 
     &+ \sum  \delta U^{(2)}_\lambda \,
    \akdag\akdag aa
    \notag\\ 
    \null &+ \sum \delta U^{(3)}_\lambda\akdag\akdag\akdag aaa
     + [\mbox{4-body}] + \cdots
    \;, 
\end{align}
where we have suppressed the single-particle indices and combinatoric factors.
As mentioned in Sec.~\ref{sec:introduction}, it is sufficient to extract the Levinger constant from the ratio of momentum distributions at high relative momentum.
We apply SRG transformations to the momentum distribution operators and use Wick's theorem in operator form to truncate at the two-body (vacuum) level.
For example, in evaluating the pair momentum distribution for two nucleons with isospin projections $\tau$ and $\tau'$ respectively, we expand and truncate
\begin{equation}
    \label{eq:n_lambda}
    \nhatlo^{\tau,\tau'}(\qvec,\Qvec) = \Uhatlam \nhathi^{\tau,\tau'}(\qvec,\Qvec) \Uhatlamdag
    ,
\end{equation}
where the unevolved ($\lambda = \infty$) pair momentum distribution operator is
\begin{equation}
    \label{eq:n_infinity}
    \nhathi^{\tau,\tau'}(\qvec,\Qvec) = 
        \frac{1}{2}\sum_{\sgn,\sgn'}
        \akdag_{\frac{\Qvec}{2} + \qvec, \sgn \tau}
        \akdag_{\frac{\Qvec}{2} - \qvec, \sgn' \tau'} 
        \ak_{\frac{\Qvec}{2} - \qvec, \sgn' \tau'}
        \ak_{\frac{\Qvec}{2} + \qvec, \sgn \tau}
        .
\end{equation}
Here $\qvec$ is the relative momentum and $\Qvec$ is the center-of-mass momentum.

At high momentum, SRG-evolved momentum distributions factorize as discussed previously.
This arises from factorization of the SRG transformation where $\delta U^{(2)}_\lambda(\kvec,\qvec) \approx \Flo(\kvec) \Fhi(\qvec)$ for $k < \lambda \ll q$.
Continuing with the pair momentum distribution as an example, at high relative momentum the operator is given by
\begin{align}
    \label{eq:pair_mom_dist_fact}
    \nhatlo(\qvec,\Qvec) &\approx \abs{\Fhi(\qvec)}^2 \sum^{\lambda}_{\kvec,\kpvec}
        \Flo(\kvec) \Flo(\kpvec) \notag \\
     & \quad\null\times   \akdag_{\frac{\Qvec}{2} + \kvec}
        \akdag_{\frac{\Qvec}{2} - \kvec} 
        \ak_{\frac{\Qvec}{2} - \kpvec}
        \ak_{\frac{\Qvec}{2} + \kpvec}
    ,
\end{align}
where we have suppressed the spin and isospin labels.

Extracting the Levinger constant involves taking a ratio of the expectation value of the operator~\eqref{eq:pair_mom_dist_fact} in a specified nucleus $A$ with the same expectation value in the deuteron.
The proton distribution operator in deuteron is given by a similar expression as Eq.~\eqref{eq:pair_mom_dist_fact} and also factorizes for $q \gg \lambda$.
In practice, we expand the $\delta U$ matrix elements in terms of partial waves and integrate over the center-of-mass momentum $Q$ in evaluating the relative pair momentum distribution.
The ratio is then dependent only on $q$ and with factorization,
\begin{equation}
    \label{eq:pn_d_ratio_factorized}
    \frac{ n_{pn}^A(q) }{ n_p^d(q) } \propto \frac{\abs{\Fhi_{pn}(q)}^2}
    {\abs{\Fhi_{d}(q)}^2} \times
    \frac{\int \mel{A}{\Flo_{pn}(k)\Flo_{pn}(k')}{A}}
    {\int \mel{d}{\Flo_{d}(k)\Flo_{d}(k')}{d}}
    ,
\end{equation}
for $q \gg \lambda$.
The soft wave functions restrict the integrals over $k$ and $k'$ to low momenta.

All partial wave channels contribute in the numerator of Eq.~\eqref{eq:pn_d_ratio_factorized} though the $\tripletS$--$\tripletD$ channel dominates (see Table~I in Ref.~\cite{Tropiano:2021qgf}).
With the denominator (deuteron) taking contributions solely from the $\tripletS$--$\tripletD$ channel, the two-body high-momentum functions $\Fhi_{pn}(q)$ and $\Fhi_{d}(q)$ roughly cancel, leaving a low-momentum ratio that is approximately scale and scheme independent, and independent of $q \gg \lambda$.
The ratio is a ``mean-field'' quantity, meaning it only depends on soft ground-state wave functions.
This is effectively the same as a ratio of GCF contacts.
We can then extract the Levinger constant from the low-momentum ratio using Eq.~\eqref{eq:gcf_levinger}.

%%%%%%%%%%%%%%%%%%%%%%%%%%%%%%%%%%%%%%%%%%%%%%%%%%%%%%%%%%%%%%%%%%%%%%%%%
\section{Results}
\label{sec:results}

\begin{figure}[tbh]
    \centering
    \includegraphics[width=0.85\columnwidth]{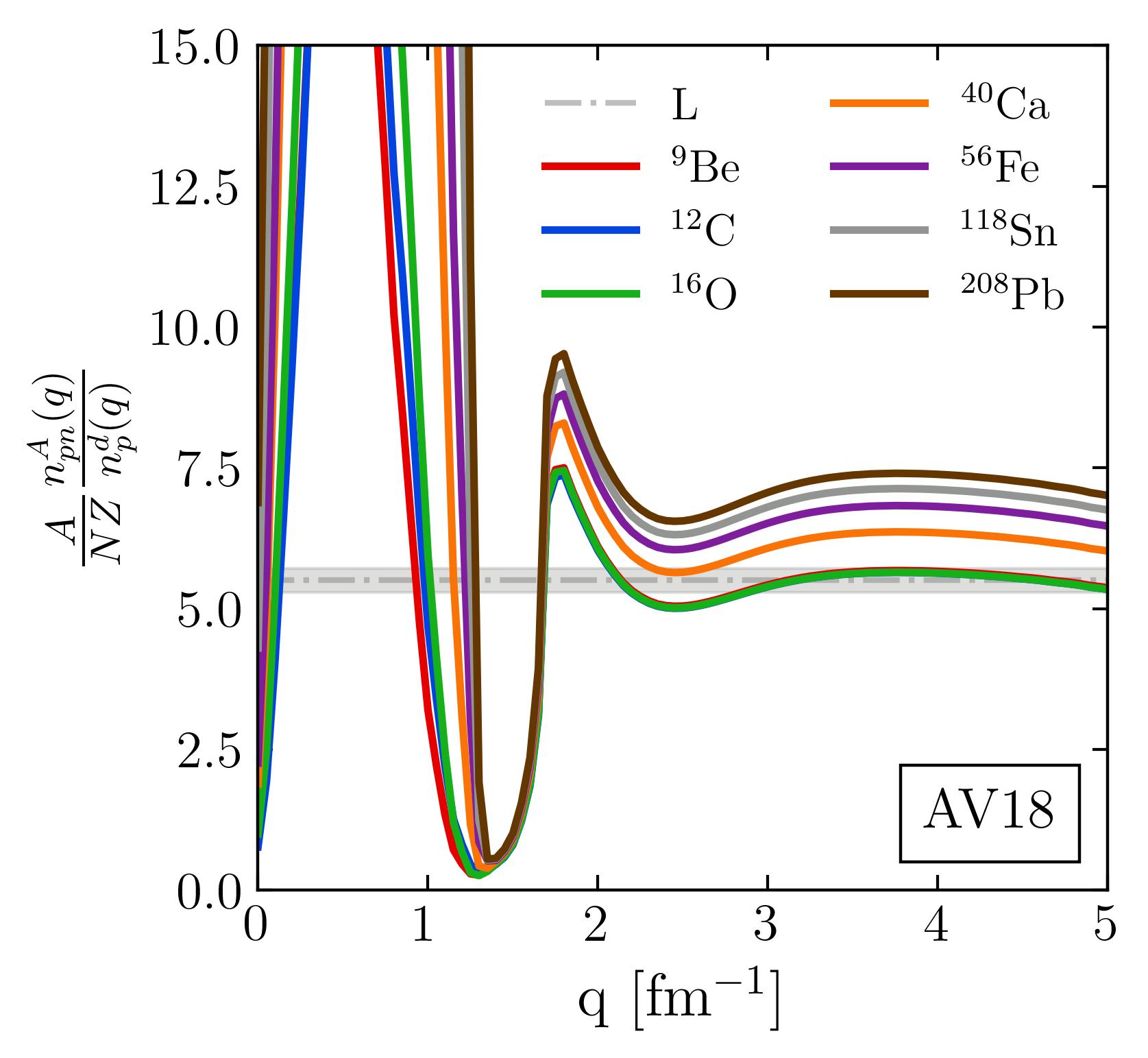}
    \caption{%
    Ratios of the $pn$ momentum distribution for nucleus $A$ over the deuteron momentum distribution as a function of relative momentum $q$ using the AV18 potential.
    The gray line and band indicate the average value of the Levinger constant with its error from Ref.~\cite{Weiss:2014gua}.
    }
    \label{fig:levinger_ratio_av18}
\end{figure}

In all results, we use densities from the Gogny functional~\cite{Decharge:1979fa} for the LDA and evolve the operators to $\lambda=1.35$\,\fmi\ including only $S$-wave contributions~\cite{Tropiano:2021qgf}.
We have made the same calculations with densities from the SLy4 Skyrme functional~\cite{Chabanat:1997un} using the HFBRAD code~\cite{Bennaceur:2005mx} and found nearly identical trends.
Inclusion of higher partial waves are not significant (as documented in Table~I of Ref.~\cite{Tropiano:2021qgf}).
We limit results to nuclei where extractions of $L$ from experimental data are available, though we can easily extend to other nuclei.

In Fig.~\ref{fig:levinger_ratio_av18} we show ratios for several nuclei of the $pn$ relative momentum distribution to the proton distribution in the deuteron, scaled by $A/NZ$.
We apply the LDA to both the numerator and denominator to help cancel systematic errors.
At high momentum the ratio plateaus to a constant value, that is, the Levinger constant.
The ratio maintains a relatively constant value across high momentum values despite the momentum distributions individually dropping several orders of magnitude.
In the low RG resolution framework, both the numerator and denominator factorize to a product of a high-momentum two-body function which carries the $q$-dependence, and a low-momentum nuclear matrix element (see Eq.~\eqref{eq:pn_d_ratio_factorized}.
The $q$-dependent functions approximately cancel in the ratio leaving a flat curve where factorization holds ($q \gg \lambda=1.35$\,\fmi) as seen in Fig.~\ref{fig:levinger_ratio_av18} for all nuclei.
The gray band shows the average value of $L=5.5$ with its uncertainty across several nuclei~\cite{Weiss:2014gua}.
The ratio tends to increase with heavier nuclei.
The behavior of the ratio near $q \sim 1.5$\,\fmi\ depends on the details of the individual momentum distributions near the Fermi surface.

\begin{figure}[tbh]
    \centering
    \includegraphics[width=0.85\columnwidth]{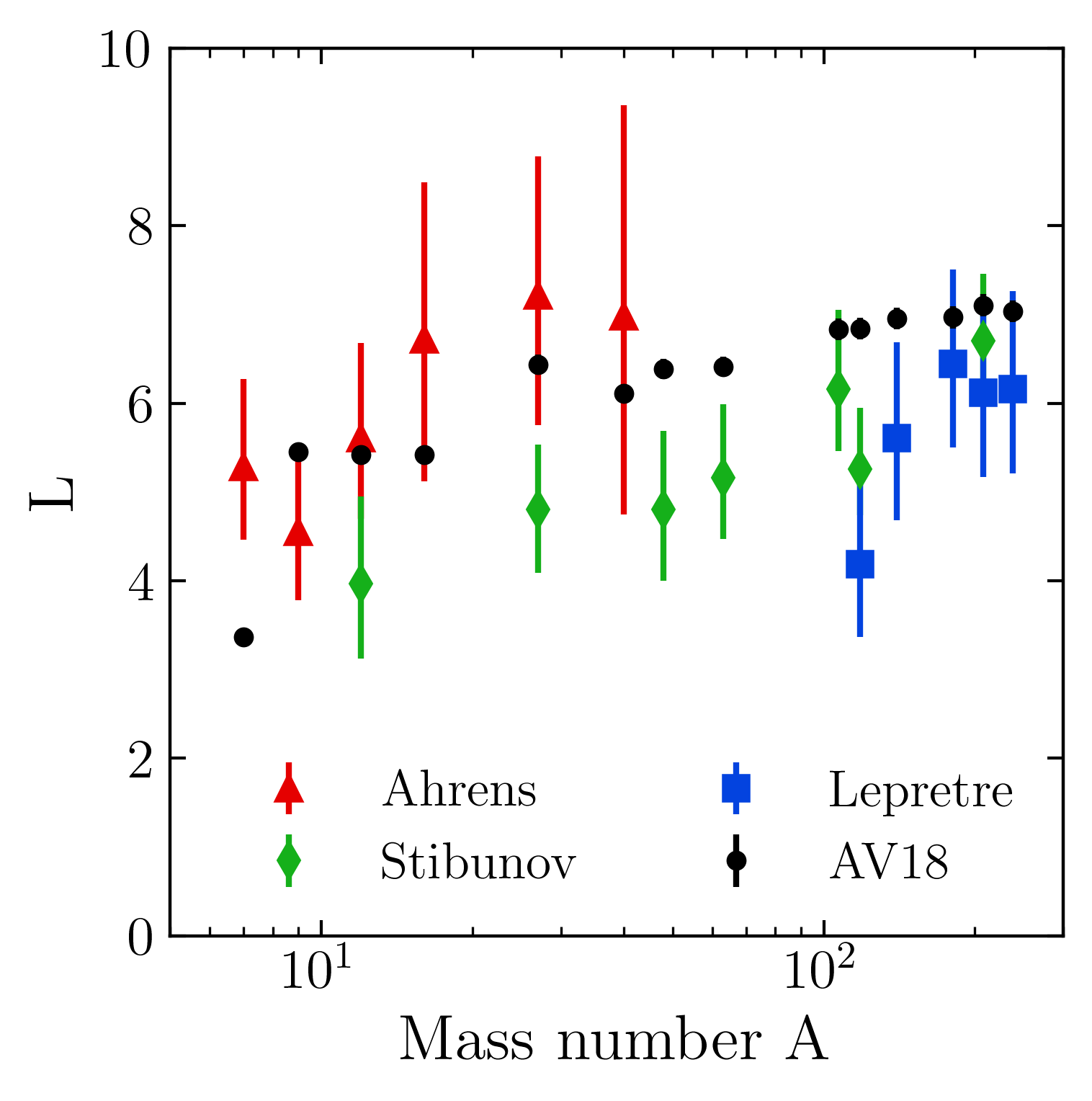}
    \caption{%
    Average Levinger constant for several nuclei with AV18 comparing to extractions from experiment.
    The error bars on the black AV18 points are determined by varying the interval of momentum $q$ in which $L$ is averaged over (see text for details).
    }
    \label{fig:levinger_constant_exp}
\end{figure}

Figure~\ref{fig:levinger_constant_exp} shows our extracted values of $L$ compared to extractions of Refs.~\cite{Terranova:1989aa,Tavares:1991zd} constrained by cross sections of nuclear photo-absorption experiments~\cite{Ahrens:1975rq,Stibunov:1984aa,Lepretre:1978me,Lepretre:1981tf}.
In extracting $L$, we take the average value of the momentum distribution ratios (shown in Fig.~\ref{fig:levinger_ratio_av18}) over $q$ from $4$ to $5$\,\fmi\ as in Ref.~\cite{Weiss:2015mba}.
Analysis with other potentials indicates that the factorization holds strongly in the momentum range of $2.5$ to $3.4$\,\fmi\ and consequently, we average over this lower momentum range as well.
Figure~\ref{fig:levinger_constant_exp} shows results from both extraction schemes where the upper (lower) error bars of the black AV18 points indicate the maximum (minimum) of the two schemes and the central value is given by the mean.
The low RG resolution calculations are in good agreement with the data and their uncertainties.
Calculated values of $L$ monotonically increase with larger mass number $A$ similar to the behavior found in calculating the SRC scaling factor $a_2$ in Ref.~\cite{Tropiano:2021qgf}.

\begin{figure}[tbh]
    \centering
    \includegraphics[width=0.85\columnwidth]{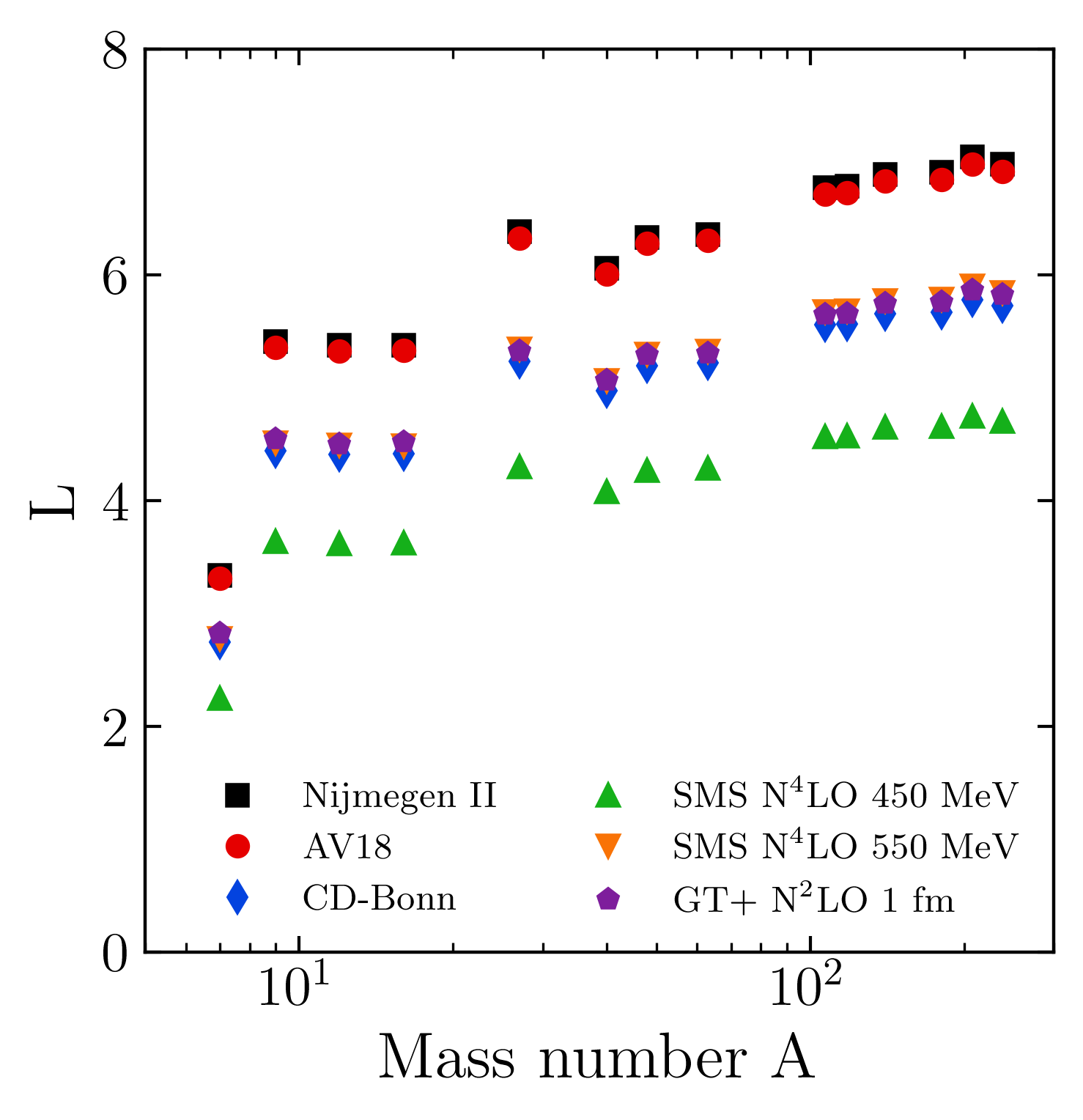}
    \caption{%
    Average Levinger constant for several nuclei comparing different NN interactions.
    }
    \label{fig:levinger_constant_kvnns}
\end{figure}

In Fig.~\ref{fig:levinger_constant_kvnns} we compare Levinger constants between several NN interactions.
We show results for AV18~\cite{Wiringa:1994wb}, Nijmegen II~\cite{Stoks:1994wp}, CD-Bonn~\cite{Machleidt:2000ge}, SMS \NNNNLO~\cite{Reinert:2017usi}, and GT+ \NNLO~\cite{Gezerlis:2014zia} averaging over the momentum range of $2.5$ to $3.4$\,\fmi.
We find that the hard potentials (e.g., AV18) produce the highest values of $L$ whereas the soft potentials (e.g., SMS \NNNNLO\ $450$\,MeV) produce relatively low values.
$L$ is extracted from the ratio of inclusive cross sections which, as an observable quantity, is RG invariant; hence, we should not find any significant discrepancies in calculations of $L$ when using different NN interactions.
However, it is incorrect to assume the same initial operator for interactions at different RG resolution scales.

\begin{figure}[tbh]
    \centering
    \includegraphics[width=0.85\columnwidth]{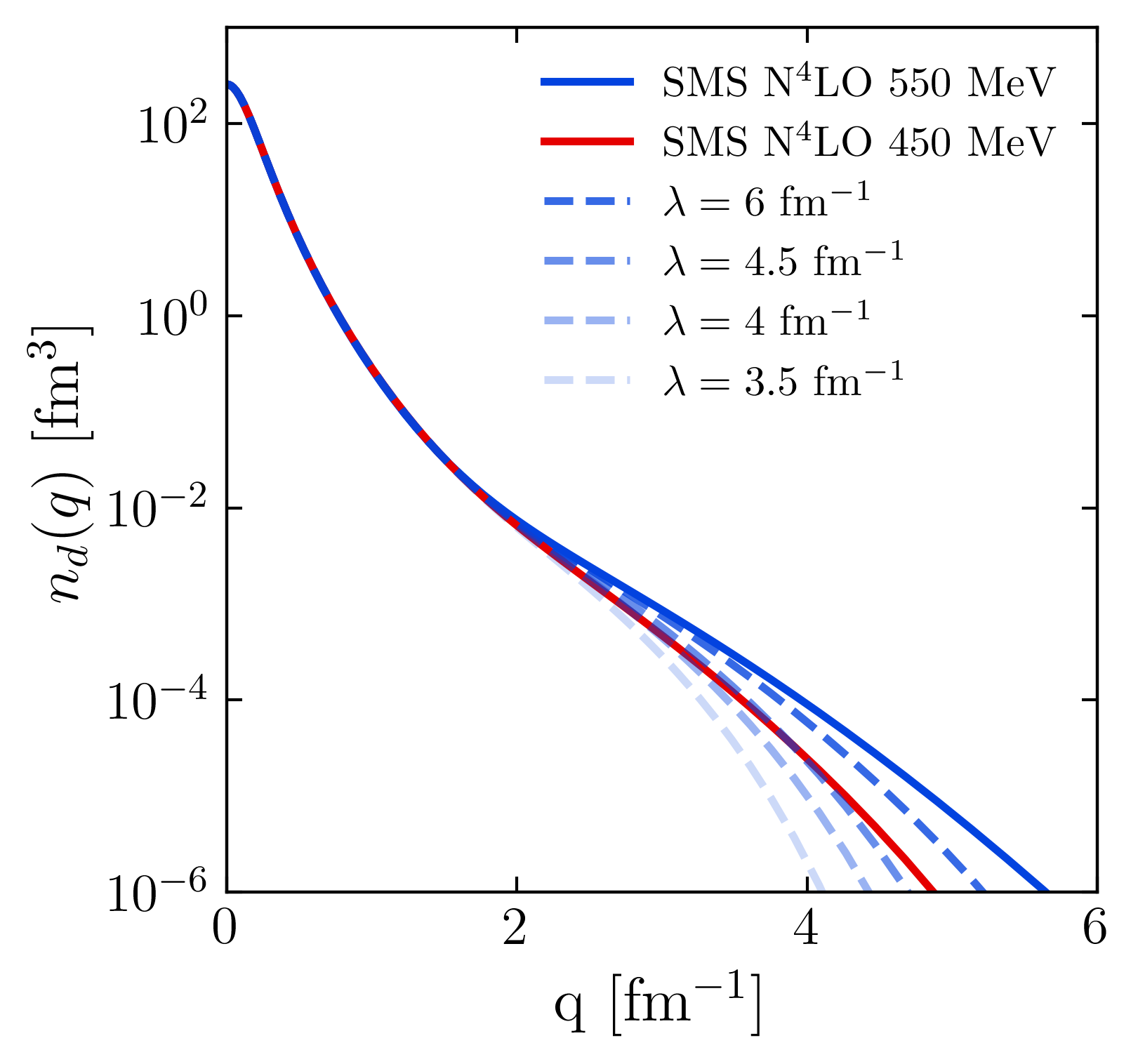}
    \caption{%
    Deuteron momentum distributions from SMS \NNNNLO\ $550$\,MeV (blue) and $450$\,MeV (red) potentials.
    The dashed lines show the distributions of the $550$\,MeV potential but SRG-evolved to some $\lambda$ value indicated by the legend.
    }
    \label{fig:sms_deuteron_distributions}
\end{figure}

The momentum distribution is resolution dependent, but we can seek to match the results of $L$ using a reference momentum distribution, in this case that of the AV18 interaction.
To do so, we must transform the momentum distribution operators of the other potentials for a consistent extraction.
For instance, if we take the usual one-body single-nucleon momentum distribution operator when using a hard potential such as AV18, then we must include an additional two-body contribution in the momentum distribution operator corresponding to the RG resolution scale of the soft potential.
In the following, we transform the initial operators of soft potentials to approximately match the results of the hard potentials using the SRG.
    
We can use SRG transformations to establish an approximate connection between two potentials of different RG resolution scales.
To illustrate this, we show the deuteron momentum distributions of the SMS \NNNNLO\ potential at two different regulator cutoffs in Fig.~\ref{fig:sms_deuteron_distributions}.
We include snapshots of the deuteron momentum distribution from the SRG-evolved ``hard'' potential ($550$\,MeV cutoff).
Around $\lambda=4-4.5$\,\fmi\ the SRG-evolved deuteron momentum distribution begins to overlap the $450$\,MeV distribution indicating a rough connection between the two potentials.
We find matching scales $\lambda \sim 3.5-5\,$\fmi\ in comparing other potentials.

We can make similar comparisons in matching interactions by considering Weinberg eigenvalues~\cite{Weinberg:1963zza}.
These eigenvalues reflect the perturbativeness of a potential and have been analyzed in several RG and EFT studies~\cite{Bogner:2006vp,Bogner:2006pc,Bogner:2005sn,Bogner:2006tw,Hoppe:2017lok}.
Consider the Born series for the $T$ matrix at energy $E$ given a Hamiltonian $H=H_0+V$,
\begin{equation}
    \label{eq:born_series}
    T(E) = V + V \frac{1}{E-H_0} V + \cdots
    .
\end{equation}
Solving for the eigenvalues and eigenvectors of the operator $(E-H_0)^{-1}V$,
\begin{equation}
    \label{eq:weinberg_eigenvalue_equation}
    \frac{1}{E-H_0} V \ket{\Gamma_{\nu}} = \eta_{\nu}(E) \ket{\Gamma_{\nu}}
    ,
\end{equation}
and applying $T(E)$ on the eigenvectors gives a power series in terms of the Weinberg eigenvalues $\eta_{\nu}(E)$,
\begin{equation}
    \label{eq:weinberg_eigenvalues}
    T(E) \ket{\Gamma_{\nu}} = (1+\eta_{\nu}(E)+\eta_{\nu}^2(E)+\cdots) V \ket{\Gamma_{\nu}}
    .
\end{equation}
Non-perturbative behavior at energy $E$ is signaled by at least one eigenvalue $|\eta_{\nu}(E)| > 1$~\cite{Weinberg:1963zza}.
For negative energies, purely attractive potentials give positive Weinberg eigenvalues and vice versa for purely repulsive potentials.
We refer to positive (negative) eigenvalues with $E\leq 0$ as attractive (repulsive); for $E>0$ the eigenvalues become complex.

\begin{table}[htb]
	\caption{%
	Largest repulsive Weinberg eigenvalues at zero energy for AV18 and SMS \NNNNLO\ $550$\,MeV evolved to several SRG resolution scales $\lambda$.
	The corresponding eigenvalue for the unevolved SMS \NNNNLO\ $450$\,MeV potential is $-0.70$.
	}
	\label{tab:weinberg_eigenvalues}
	\begin{ruledtabular}
		\begin{tabular}{c|cccc}
		   $\lambda$ [\fmi] & AV18 & \null & $550$\,MeV & \null \mystrut\\
		    \colrule
      			$\infty$ & $-3.06$ & \null & $-1.22$ & \null \mystrut\\
      			$12$ & $-2.94$ & \null & $-1.27$ & \null \mystrut\\
      			$6$ & $-1.81$ & \null & $-1.10$ & \null \mystrut\\
      			$5.5$ & $-1.61$ & \null & $-1.05$ & \null \mystrut\\
      			$5$ & $-1.40$ & \null & $-0.98$ & \null \mystrut\\
      			$4.5$ & $-1.19$ & \null & $-0.88$ & \null \mystrut\\
      			$4$ & $-0.98$ & \null & $-0.78$ & \null \mystrut\\
      			$3.5$ & $-0.79$ & \null & $-0.66$ & \null \mystrut\\
      			$3$ & $-0.62$ & \null & $-0.54$ & \null \mystrut
		\end{tabular}
  	\end{ruledtabular}
\end{table}

In comparing interactions, we evolve the harder of the two potentials, compute the Weinberg eigenvalues at zero energy, and do the same for several SRG-evolved versions of the same potential.
Then we compute the corresponding Weinberg eigenvalues of the softer potential and compare to the largest repulsive eigenvalues to determine the matching scale.
We document our results for the Weinberg eigenvalues in Table~\ref{tab:weinberg_eigenvalues}.
This analysis gives roughly the same matching scales as found with comparing deuteron momentum distributions.

\begin{figure}[tbh]
    \centering
    \includegraphics[width=0.85\columnwidth]{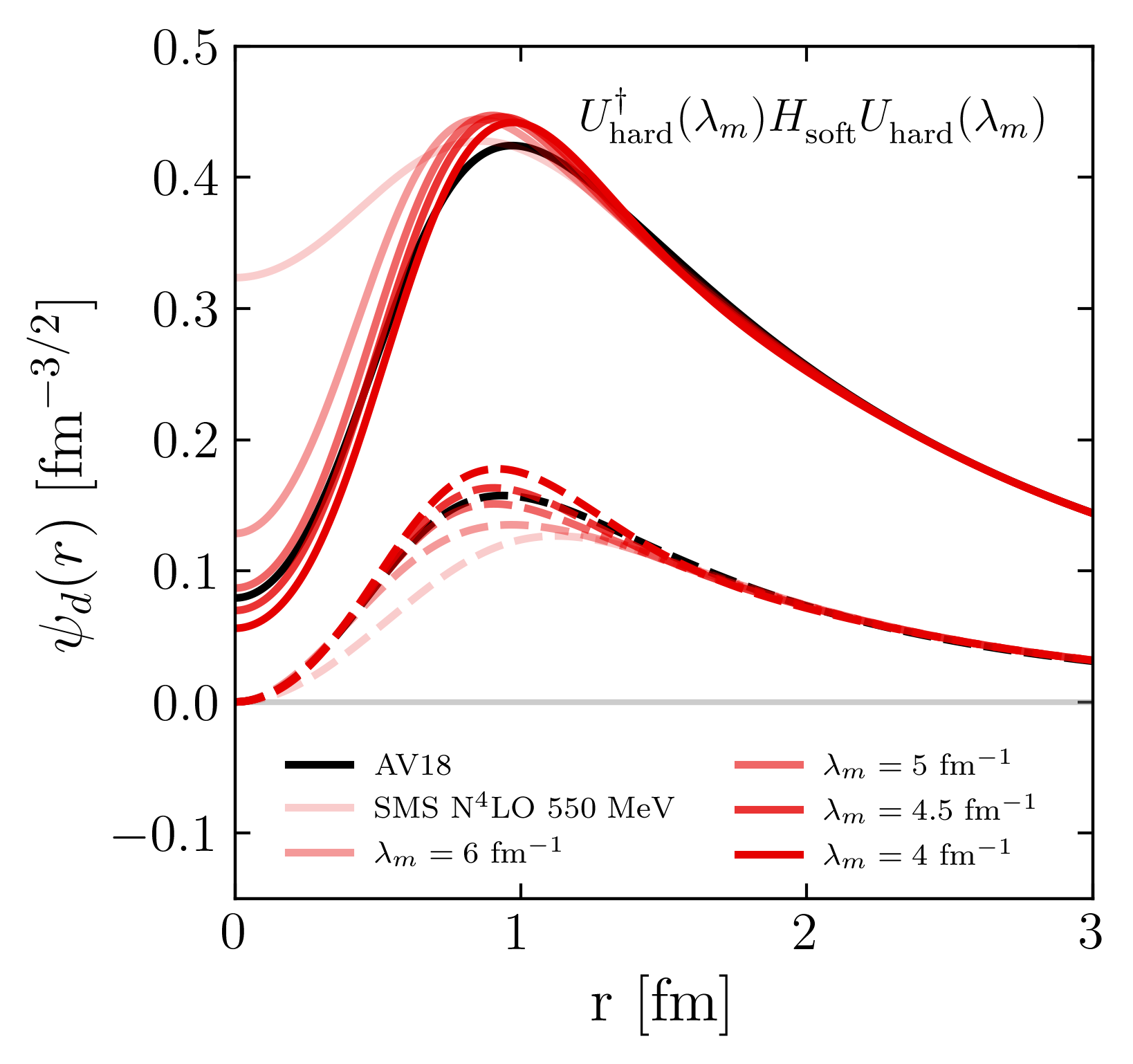}
    \caption{%
    Deuteron wave functions of AV18 (black) and SMS \NNNNLO\ $550$\,MeV (red) in coordinate space.
    Additionally, we show deuteron wave functions of SMS \NNNNLO\ $550$\,MeV but inverse-SRG transformed with AV18 at several $\lambda_m$ values.
    The solid lines correspond to the S states, and the dashed lines correspond to the D states.
    }
    \label{fig:sms_av18_deuteron_wfs}
\end{figure}

In the following, we extract the Levinger constant matching two potentials of different RG resolution scales.
We use a scale denoted $\lambda_m$ associated with the matching scale from the previous analysis to apply inverse-SRG transformations of the harder potential onto the softer of the two potentials.
Figure~\ref{fig:sms_av18_deuteron_wfs} compares deuteron wave functions of AV18 and SMS \NNNNLO\ $550$\,MeV, including several inverse-SRG evolved snapshots of the SMS wave function.
These are inverse-SRG transformations from the AV18 interaction; hence, the deuteron wave function gains a \emph{stronger} short-range modification as $\lambda_m$ decreases.
At $r=0$ the inverse-evolved deuteron wave function matches AV18 for $\lambda_m$ in between $5$ and $4.5$\,\fmi, in agreement with the scale found from analyzing Weinberg eigenvalues.

\begin{figure}[tbh]
    \centering
    \includegraphics[width=0.85\columnwidth]{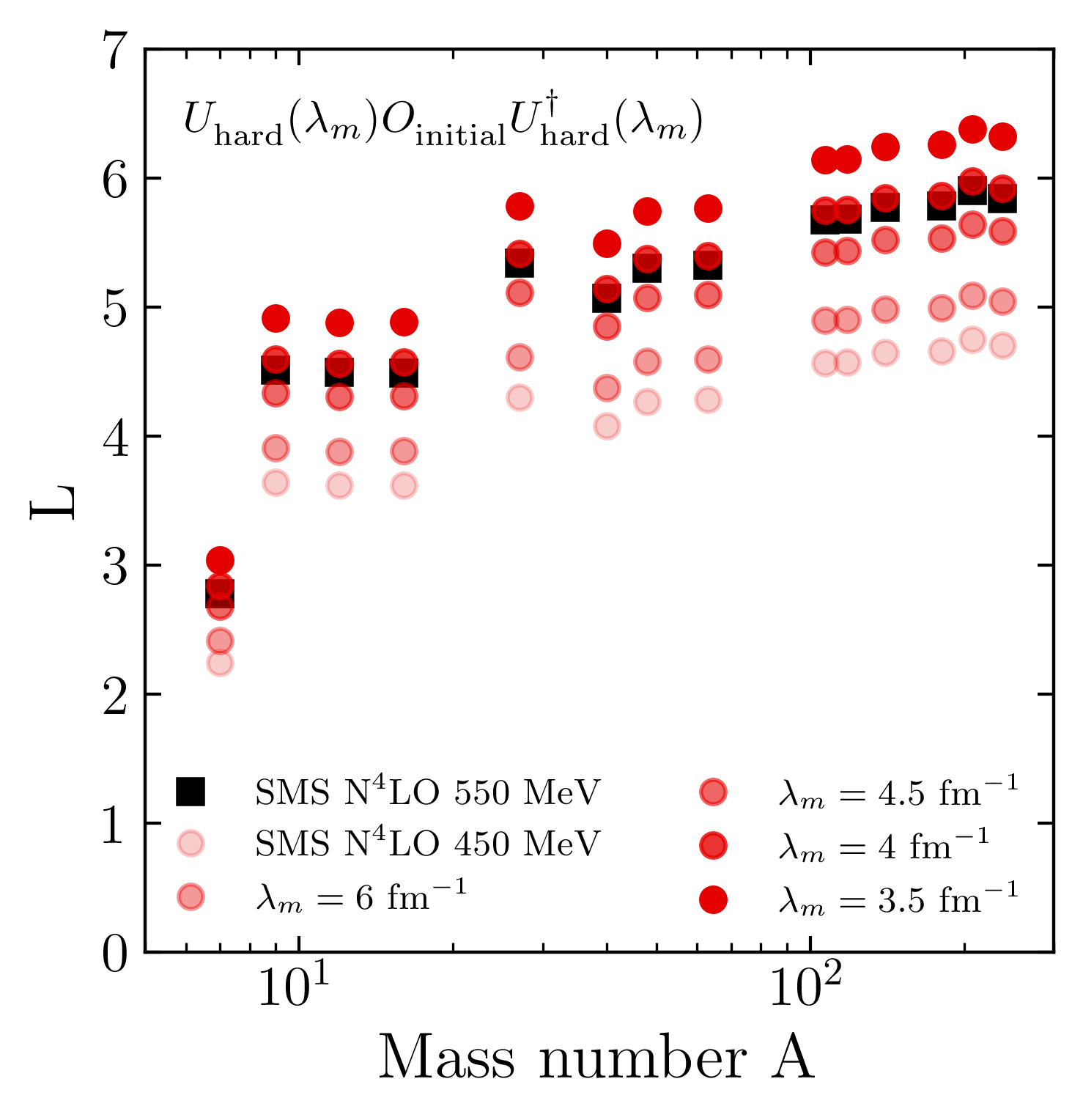}
    \caption{%
    Average Levinger constant for several nuclei comparing the SMS \NNNNLO\ $550$\,MeV (black) and $450$\,MeV (red) potentials.
    Results are also shown for the SMS \NNNNLO\ $450$\,MeV potential with an additional two-body operator due to inverse-SRG transformations from SMS \NNNNLO\ $550$\,MeV at several values of $\lambda_m$.
    }
    \label{fig:levinger_constant_scale_dependence_sms}
\end{figure}
\begin{figure}[tbh]
    \centering
    \includegraphics[width=0.85\columnwidth]{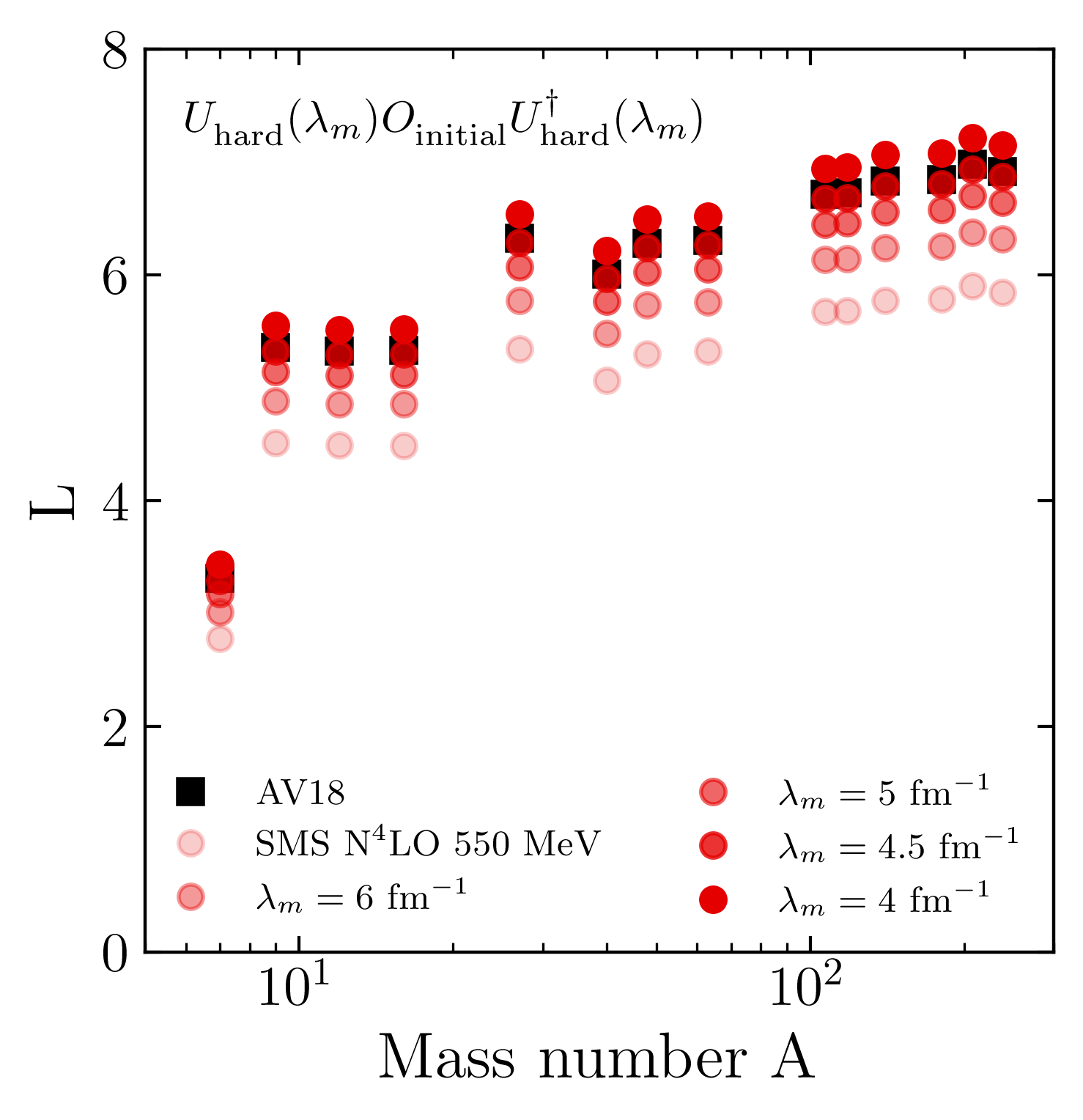}
    \caption{%
    Same as Fig.~\ref{fig:levinger_constant_scale_dependence_sms} but comparing SMS \NNNNLO\ $550$\,MeV to AV18.
    }
    \label{fig:levinger_constant_scale_dependence_sms_av18}
\end{figure}

Taking the inverse-SRG-evolved, soft Hamiltonian as the initial Hamiltonian, then evolving that Hamiltonian down to $\lambda=1.35$\,\fmi\ in calculating the momentum distribution, is equivalent to applying transformations of the hard potential on the initial operator
\begin{equation}
    \label{eq:soft_initial_operator}
    \Ohat_{\lambda_m} = \Uhat^{\phantom{\dagger}}_{\lambda_m} \Ohat \Uhatdag_{\lambda_m}
    .
\end{equation}
Here $\Ohat$ is the initial operator to be used with the hard potential, $\Uhat_{\lambda_m}$ corresponds to transformations of the hard potential ($\lambda_m \approx 4.5$\,\fmi), and $\Ohat_{\lambda_m}$ is the initial operator to be used with the soft potential.
Consequently, the soft potentials start with an additional two-body contribution in the operator, whereas hard potentials (such as AV18) start with solely the momentum projection operator~\eqref{eq:n_infinity}.
Figure~\ref{fig:levinger_constant_scale_dependence_sms} compares the two SMS \NNNNLO\ potentials showing results for several $\lambda_m$ values.
We see that the Levinger constants computed from the $450$\,MeV potential are raised to match the values from the $550$\,MeV potential around $\lambda_m=4-4.5$\,\fmi, consistent with the deuteron and Weinberg eigenvalue comparisons.
These results confirm that an additional two-body operator is necessary in calculating consistent values of $L$ for potentials of a low RG resolution.
    
In Fig.~\ref{fig:levinger_constant_scale_dependence_sms_av18} we make the same comparison as in Fig.~\ref{fig:levinger_constant_scale_dependence_sms} but for AV18 and SMS \NNNNLO\ $550$\,MeV.
Here we see that $\lambda_m \approx 4.5$\,\fmi\ gives the matching scale between the two potentials.
Note, this method only serves as an approximate tool in matching interactions, but in general there will be additional differences in comparing interactions due to scheme dependence (regulator differences, coordinate- or momentum-space formulations, and so on.)

%%%%%%%%%%%%%%%%%%%%%%%%%%%%%%%%%%%%%%%%%%%%%%%%%%%%%%%%%%%%%%%%%%%%%%%%%
\section{Summary and outlook}
\label{sec:summary}

We have shown how the Levinger constant can be quantitatively calculated at low RG resolution with simple approximations.
This analysis relied on our previous work in using LDA estimates to calculate evolved momentum distributions.
The observed scale (and scheme) dependence of the extracted Levinger constants reflects in part insufficient matching of the reaction operator, either to experiment or a more accurate [high resolution] theoretical description.
Additional two-body contributions induced by inverse-SRG transformations on the initial operator restores approximate scale independence.

This strategy demonstrates a more general concept: NN interactions can be ``smoothly'' connected by RG transformations.
The matching can be done by comparing deuteron wave functions or Weinberg eigenvalues, with consistent results for the matching scale.
It may be adequate to incorporate only a contact interaction instead of applying inverse-SRG transformations to match the initial operator to its associated interaction.
We leave this point as a follow-up for future work.

The path to more precise determinations is clear.
In particular, there are several classes of corrections that need to be examined going forward:
\begin{itemize}
    \item Incorporating improved many-body physics to test the limits of LDAs and enable uncertainty quantification.
    The LDA approximation here is implemented as the leading-term in a density matrix expansion~\cite{Negele:1972zp}.
    Including next-to-leading terms is the next step.
    \item Solving the many-body problem with SRG-evolved operators at different $\lambda$ values allows for an indirect method in estimating three-body contributions.
    Residual $\lambda$ dependence indicates the size of neglected contributions from induced three-body and higher-body terms~\cite{Jurgenson:2009qs,Neff:2015xda}.
    However, computational restrictions may limit this approach to light nuclei.
    \item Understanding the impact of long-range correlations and isolating from short-range correlations.
    \item Better understanding and exploiting the SRG resolution ($\lambda$) dependence (e.g., optimal value of $\lambda$ given our approximations).
\end{itemize}
To better quantify the impact of common approximations we will directly calculate photo-absorption cross sections, following the low RG resolution calculation of deuteron electrodisintegration~\cite{More:2015tpa,More:2017syr}.

A natural follow-up to the present work is a general examination of level depletion in the RG framework.
The results here and in Ref.~\cite{Tropiano:2021qgf} have explicitly established how processes with particular types of high-energy final states are directly accounted for at low RG resolution by the evolution of basic reaction operators.
The converse task is to quantitatively understand at low RG resolution the physics associated at high resolution with depletion of single-particle states~\cite{Tostevin:2014usa,Aumann:2020tcq}, given that these states are largely occupied in the mean-field picture at low resolution.
Here the role of mid-to-long-range correlations are particularly important to understand.
This task will involve extending the low RG resolution framework to knockout reactions with electron or nucleon probes.
In either case optical potentials play an important role in modeling these processes, and consequently, we must understand how optical potentials change under RG evolution, which
will build on Ref.~\cite{Hisham:2022prep}.

%\vspace*{.01in}
%%%%%%%%%%%%%%%%%%%%%%%%%%%%%%%%%%%%%%%%%%%%%%%%%%%%%%%%%%%%%%%%%%%%%%%%%
\begin{acknowledgments}
We thank Alberto Garcia for fruitful discussions and feedback, and Nicolas Schunck for sharing a code to calculate Gogny nucleon densities.
The work of AJT, RJF, and MAH was supported by the National Science Foundation under Grant No.~PHY--1913069, and the NUCLEI SciDAC Collaboration under US Department of Energy MSU subcontract RC107839-OSU\@.
The work of SKB was supported by the National Science Foundation under Grant Nos. PHY-1713901 and PHY-2013047, and the U.S. Department of Energy, Office of Science, Office of Nuclear Physics under Grant No. de-sc0018083 (NUCLEI SciDAC Collaboration).
\end{acknowledgments}

\bibliography{tropiano_bib}

\end{document}